\documentclass[twocolumn,amsmath,showpacs,amsfonts,aps,prc,floatfix]{revtex4}
\usepackage{graphicx}
\usepackage{bm}
\newcommand{\la}{\langle}
\newcommand{\ra}{\rangle}

\newcommand{\vih}{\left (\frac{v_2}{\epsilon} \right )^{ih}}
\newcommand{\etas}{\left ( \frac{1}{\tau_i T_i} \frac{\eta}{s} \right )}
\newcommand{\vt}{v_2 \{2\}}
\newcommand{\vf}{v_2 \{4\}}

\begin{document}
\title{Modified Knudsen ansatz and elliptic flow in $\sqrt{s}$=14 TeV pp collisions}
 
\author{A. K. Chaudhuri}
\email[E-mail:]{akc@veccal.ernet.in}
\affiliation{Variable Energy Cyclotron Centre, 1/AF, Bidhan Nagar, 
Kolkata 700~064, India}

\begin{abstract}

Assuming that hot spots are formed in initial pp collisions, in a modified Knudsen ansatz, which accounts for the entropy generation in viscous fluid evolution, we have given predictions for elliptic flow in
  $\sqrt{s}$=14 TeV pp collisions. Predicted flow depends on the number of hot spots and hot spot sizes. If two to four hot spots of size $\approx$0.1 fm are formed in initial pp collisions, in events with multiplicity $n_ {mult}\approx$10-15,  modified Knudsen ansatz predicted flow is accessible experimentally in 4th order cumulant method.
\end{abstract}

\pacs{25.75.-q, 25.75.Dw, 25.75.Ld} 

\date{\today}  

\maketitle


 In recent years, there is much interest in elliptic flow in pp collisions at LHC energy. Finite elliptic flow has been observed in $\sqrt{s}_{NN}$=200 GeV Au+Au collisions \cite{BRAHMSwhitepaper,PHOBOSwhitepaper,PHENIXwhitepaper,STARwhitepaper} and more recently in $\sqrt{s}_{NN}$=2.76 TeV Pb+Pb collisions \cite{Aamodt:2010pa}. 
Finite elliptic flow in relativistic heavy ion collisions is regarded as a definitive signature of collective effect  \cite{Ollitrault:1992bk,Poskanzer:1998yz}. It is also best understood in a collective model like hydrodynamics \cite{QGP3}. In a non-central collision, the reaction zone is spatially asymmetric. Differential pressure gradient convert the spatial asymmetry in to momentum asymmetry. In other words, in a hydrodynamic model, spatial asymmetry ($\varepsilon_x=\frac{<y^2>-<x^2>}{<y^2>+<x^2>}$) of the interaction region controls the elliptic flow.  
Since protons have finite extension (though of smaller size than a nucleus), in principle, in finite impact parameter pp collisions asymmetric reaction zone 
can produce elliptic flow. 
Similarities between pp and Au+Au collisions have been observed even at RHIC energy \cite{Chajecki:2009es}.
When phase space restriction due to conservation laws is taken into account
transverse momentum distribution in pp and Au+Au collisions at RHIC energy  show similar behavior \cite{Chajecki:2009es}. However, similarity in $p_T$ spectra alone does not prove that collective model like hydrodynamic is applicable in pp collisions. Observation of finite elliptic flow could be a definitive signature of collective behavior in pp collisions. However, even if flow is produced in pp collisions, whether or not it will be accessible experimentally will depend 
 on both the flow strength and the multiplicity in the phase space window where the flow is measured.
This is because non-flow effects like di-jet production, also show azimuthal correlation not related to the reaction plane. They need to be disentangled for faithful reconstruction of the reaction plane. Several standard methods  \cite{Poskanzer:1998yz,Ollitrault:1993ba,Borghini:2001vi,Bhalerao:2003xf} have been devised to discriminate non-flow effects. Event plane method \cite{Poskanzer:1998yz,Ollitrault:1993ba} determine the reaction plane, but require large multiplicity for unambiguous determination. 
Cumulant method \cite{Borghini:2001vi} does not require measurement of the reaction plane. Cumulants of multiparticle azimuthal correlation are related to flow harmonics.  The cumulants can be constructed in increasing order according to the number of particles that are azimuthally correlated. The method relies on the different 
multiplicity scaling property of the azimuthal correlation related to flow and non-flow effects. In the cumulant method, for particle multiplicity $n_{mult}$, $v_2$ can be reliably extracted using two particle correlator, if $\vt > 1/n_{mult}^{1/2}$. Higher order correlators will increase the sensitivity, e.g.
$\vf > 1/n_{mult}^{3/4}$. 
Still higher order cumulant (cumulants of order greater than 4) will increase the sensitivity even more, $v_2   >1/n_{mult}$.  In the Lee-Yang zero method \cite{Bhalerao:2003xf} elliptic flow is obtained from
the zeros in a complex plane of a generating function of azimuthal correlation.
It is also less biased by the non-flow correction, $v_2\{ Lee-Yang \}>1/n_{mult}$. 

Multiplicity in a pp collision is not large. For example, in the central rapidity region, $|\eta| < 1$, in $\sqrt{s}$=7 TeV pp collisions, ALICE collaboration measured charged particle density  $dN_{ch}/d\eta \approx6$ \cite{Aamodt:2010pp}.  If for every charged pair, there is a neutral particle, 
$n_{mult}\approx$ 9 in $\sqrt{s}$=7 TeV pp collisions.  
Unless the elliptic flow $v_2 > 1/n^{3/4}_{mult} \approx 0.2$, experimentally flow can not be measured   in the 4th order   cumulant method. 
In $\sqrt{s}$=14 TeV pp collisions, multiplicity is expected to increase. Extrapolation to existing ALICE data gives $n_{mult}\approx $11 in $\sqrt{s}$=14 TeV pp collisions. Only $v_2 \geq 0.16$,  can possibly be measured. Note that $v_2 \approx$ 0.16-.2 is a very large value.
For example, in $\sqrt{s}$=2.76 TeV Pb+Pb collisions, in a peripheral 30-40\% collision, in the central rapidity region, multiplicity is $\sim$ 640 and elliptic flow $v_2\sim$0.08 \cite{Aamodt:2010pa}.   Recently, in \cite{Prasad:2009bx}, $\sqrt{s}$=14 TeV pp collisions were simulated hydrodynamically. The model parameters like initial time, initial energy density distribution, freeze-out temperature were fixed to reproduce expected charged particles  multiplicity ($dN/dy\approx 7$) in a minimum bias collision. 
Elliptic flow as a function of centrality was studied. Even in a peripheral collision, hydrodynamic predictions for $v_2$ is small, $v_2< 0.02$.
Unless some exotic mechanism is at work, it is unlikely that $v_2\approx 0.16-0.2$ can be generated in pp collisions. 

Recently some authors have considered exotic mechanism  like hot spot formation in pp collisions at LHC \cite{CasalderreySolana:2009uk},\cite{Bozek:2009dt},\cite{Chaudhuri:2009yp}.
Elliptic flow is proportional to initial spatial eccentricity. With hot spots, even in a central collision, spatial eccentricity becomes non-zero and measurable elliptic flow can be generated. 
In \cite{CasalderreySolana:2009uk}, it was argued that if 2-3 hot spots are formed in pp collisions, in events with multiplicity $n_{mult} >$50, experimentally measurable flow can be generated. The  widely known Knudsen ansatz \cite{Bhalerao:2005mm}, 

\begin{equation} \label{eq1}
\left (\frac{v_2}{\epsilon} \right )^{ex}=\left (\frac{v_2}{\epsilon} \right )^{ih} 
\frac{  \sigma c_s \frac{1}{S}\frac{dN}{dy}    }
{\frac{1}{K_0} + \sigma c_s \frac{1}{S}\frac{dN}{dy},   
}
\end{equation} 

\noindent was used to obtain the estimate of elliptic flow in pp collisions.  
In Eq.\ref{eq1}, $\vih$ is the hydrodynamic limit for the elliptic flow, $S$ is the transverse area of the reaction zone, $\sigma$ is the interparticle cross section, $c_s$ is the speed of sound of the medium and $K_0$ is a non-linear parameter of order $\sim$ 1, whose exact value can be obtained from explicit transport calculation \cite{Gombeaud:2007ub}. 
  $\sigma c_s \frac{1}{S}\frac{dN}{dy}$ can be identified with the inverse Knudsen number, $K^{-1}=\sigma c_s \frac{1}{S}\frac{dN}{dy}$. Eq.\ref{eq1} give qualitatively correct behavior of the experimental elliptic flow. 
 In the limit of small Knudsen number experimental flow approach the ideal hydrodynamic limit $\vih$ with a small correction. In the other extreme limit of large Knudsen number, flow is proportional to Knudsen number.

 \begin{figure}[t]
\center
 \resizebox{0.35\textwidth}{!}{%
  \includegraphics{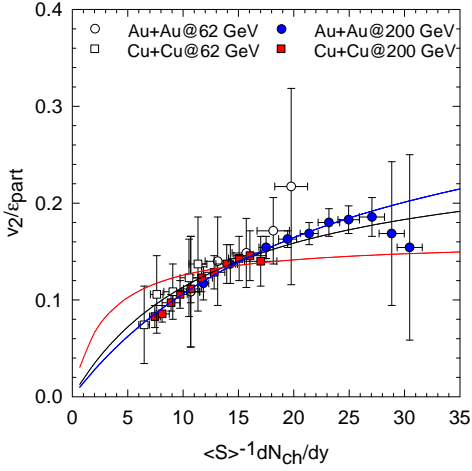}
}
\caption{(color online) PHOBOS data for the centrality dependence of eccentricity scaled elliptic flow in Cu+Cu and Au+Au  collisions at $\sqrt{s}$= 62 and 200 GeV. The black and red   lines are the fit in the unmodified Knudsen ansatz (ideal fluid) with $K_0\sigma c_s$=0.05 and 0.23 respectively. The blue solid and dashed line are the fit in the modified Knudsen ansatz (viscous fluid) with $K_0\sigma c_s$=0.05 and 0.23 respectively.} \label{F1}
\end{figure}

However, Eq.\ref{eq1} 
is valid only in the ideal fluid approximation. 
It was obtained with the assumption that the total particle number is conserved throughout the evolution \cite{Bhalerao:2005mm}. The assumption is
justified in an isentropic expansion, i.e. one dimensional evolution of ideal fluid, when entropy density ($s$) times the proper time ($\tau$) is a constant. Under such condition,   $\frac{1}{S} \frac{dN}{dy} \propto s\tau \approx n \tau$ \cite{Hwa:1985xg}. However, in a viscous evolution, entropy is generated and initial and final state entropy are not same and the assumption is clearly violated. In \cite{Chaudhuri:2010in}, Eq.\ref{eq1} was extended to include the effect of entropy generation. In the modified Knudsen ansatz,
  
\begin{equation} \label{eq2}
\left (\frac{v_2}{\epsilon} \right )^{ex}=\left (\frac{v_2}{\epsilon} \right )^{ih} 
\frac{  \frac{1}{S}\frac{dN}{dy} \left [1+\frac{2}{3\tau_iT_i} \left(\frac{\eta}{s}\right)  \right ]^{-3}  }
{\frac{1}{K_0 \sigma c_s} + \frac{1}{S}\frac{dN}{dy} \left [1+ \frac{2}{3\tau_iT_i} \left(\frac{\eta}{s}\right)  \right ]^{-3}  
}
\end{equation} 

\begin{figure}[t]
\vspace{0.35cm} 
\center
 \resizebox{0.3\textwidth}{!}{%
  \includegraphics{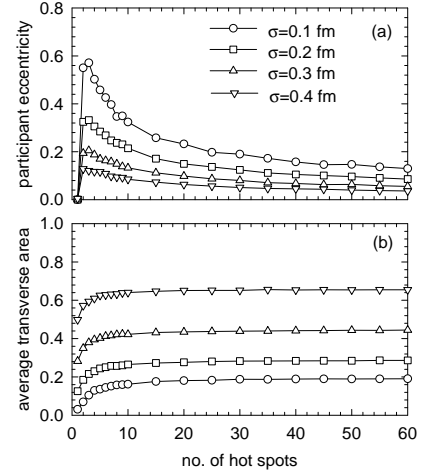} 
}
\caption{ (a) Variation of event averaged participant eccentricity with hot spot number and size. (b) same for the average transverse area. } \label{F2}
\end{figure}

\noindent where $\eta/s$ is the viscosity to entropy ratio of the medium. $\tau_i$ and $T_i$ is the initial time and temperature scale.  
The modified equation clearly brought out the effect of viscosity on elliptic flow. Inverse Knudsen number is reduced, $K^{-1}=\sigma c_s \frac{1}{S}\frac{dN}{dy} \rightarrow \sigma c_s\frac{1}{S}\frac{dN}{dy}[1+\frac{2}{3}\etas]^{-3}$. To reproduce the experimental flow, reduction in $K^{-1}$ must be compensated by increase in ideal hydrodynamic limit $\vih$. This is an interesting result. Modified Knudsen ansatz  require more $\vih$ than the unmodified one.

The modified Knudsen ansatz do explains the experimentally observed centrality dependence of elliptic flow. As an example, in Fig.\ref{F1}, fits obtained to the PHOBOS measurements \cite{Back:2004mh,Alver:2006wh,Alver:2008zza} for charged particles elliptic flow in $\sqrt{s}$=62 and 200 GeV Au+Au and Cu+Cu collisions are shown. Within the error, PHOBOS measurements of (participant) eccentricity scaled elliptic flow do not show any energy dependence or system size dependence. In the modified Knudsen ansatz, treating $\vih$ and $\etas$ as free parameters, we have fitted the PHOBOS data. In the Knudsen ansatz, the combined parameter $K_0\sigma c_s$ needs to be specified. We have used two  values, $K_0\sigma c_s$=0.05 and 0.23 to account for the uncertainty in the parameters $c_s=\sqrt{1/3}$, $\sigma$=3-4 mb and $K_0=0.7 \pm 0.3$ \cite{Gombeaud:2007ub}. The fitted values, along with the $\chi^2/N$ of the fit, are listed in table.\ref{table1}. The    dashed and solid blue lines in Fig.\ref{F1} are the fits obtained with $K_0\sigma c_s$=0.05 and 0.23 respectively. The two fits can not be distinguished. Ideal hydrodynamic limit $\vih$=0.36 is  also identical for both the values of $K_0\sigma c_s$. The fitted value of $\etas$ however differ by a factor of $\sim$5, $\etas$=0.29 for $K_0\sigma c_s$=0.05 and $\etas$=1.37 for $K_0\sigma c_s$=0.23. Increase of $\etas$ with $K_0\sigma c_s$ is also understood. From Eq.\ref{eq2}, one immediately gets, $\left (\frac{v_2}{\epsilon} \right ) ^{ex}\propto \frac{K_0\sigma c_s}{\etas}$. Factor of $\sim$5 increase in $K_0\sigma c_s$  is compensated by similar increase in $\etas$. 

For comparison, in Fig.\ref{F1}, fits obtained to the PHOBOS data  in the unmodified Knudsen ansatz, i.e. without accounting for the entropy generation are also shown. The black and red lines in Fig.\ref{F1} are the fits obtained in the unmodified Knudsen ansatz respectively for $K_0\sigma c_s$=0.05 and 0.23.   The fitted value of $\vih$ are listed in table.\ref{table1}. For both the values of $K_0\sigma c_s$, ideal hydrodynamic limit $\vih$ is less than that obtained in the modified Knudsen ansatz.

 \begin{figure}[t]
\center
 \resizebox{0.35\textwidth}{!}{%
  \includegraphics{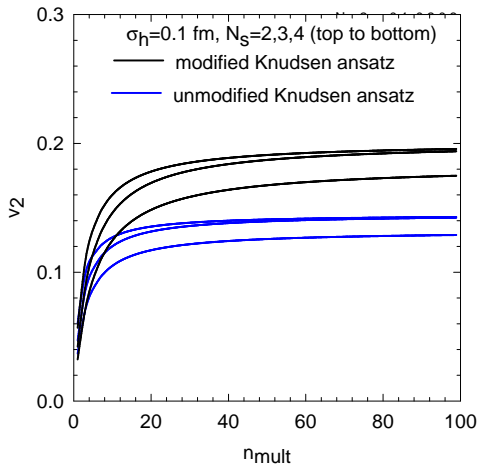}
}
\caption{(color online) Black and blue lines (from top to bottom) are predicted flow in modified and unmodified  Knudsen ansatz with $N_s$=2, 3 and 4 hot spots of size 0.1 fm.  
}
 \label{F3}
\end{figure}

If the participant eccentricity and the transverse area are known, 
results of the analysis of PHOBOS data can be used to predict for elliptic flow in pp  collisions. We assume that hot spots are formed in initial pp collisions. They   have Gaussian density distribution. For $N_s$ number of hot spots, the energy density of the system can be obtained as,

\begin{equation}
\varepsilon({x,y})=\varepsilon_0 \frac{1}{\sqrt{2 \pi \sigma^2}}
\sum_{i=1}^{N_s} e^{-\frac{({\bf r}-{\bf r}_i)^2}{2\sigma_h^2}}
\end{equation}

The centre of the hot spots (${\bf r}_i$) can be anywhere in the reaction volume. We assume that ${\bf r}_i$'s are 
randomly distributed within a  sphere of radius R=0.56 fm. 
Spatial eccentricity of the reaction zone will depend on the number of hot spots as well as on the size $\sigma_h$ of the hot spots. In Fig.\ref{F2}, variation average participant eccentricity 
($\langle \varepsilon_{part} \rangle$) and transverse area $S$, with number of hot spots as well as with the size $\sigma_h$ is shown. To be consistent with PHOBOS measurements,   $\langle \varepsilon_{part} \rangle$ and $S$ are computed as follows:

\begin{eqnarray}\label{eq4}
\la \varepsilon_{part} \ra&=&
\frac{\sqrt{(\sigma_y^2-\sigma_x^2)+4\sigma_{xy}^2}}{\sigma_y^2+\sigma_x^2}\\
 S  &=&\pi \sqrt{\sigma^2_x \sigma^2_y-\sigma^2_{xy}}
\end{eqnarray}

\noindent where $\sigma_x^2=\la x^2 \ra-\la x\ra^2$,$\sigma_y^2=\la y^2 \ra-\la y\ra^2$ and $\sigma_{xy}=\la xy \ra-\la xy\ra$, and 
$\la ... \ra$ denote energy density weighted averaging.

If a single hot spot is formed, the  $\langle \varepsilon_{part} \rangle \approx 0$. $\langle \varepsilon_{part} \rangle$ is maximum if only two hot spots are formed in the initial collisions.   $\langle \varepsilon_{part}\rangle$ decreases
as more and more hot spots are formed and for very large number of hot spots $\langle\varepsilon_{part} \rangle \rightarrow$ 0.  $\langle \varepsilon_{part}\rangle$ also decreases with increasing size of the hot spots.    Average transverse area ($S$) increases with $N_s$ and saturates beyond $N_s\approx$5. $S$ also increases with the hot spot size.

\begin{figure}[t]
\center
 \resizebox{0.35\textwidth}{!}{%
  \includegraphics{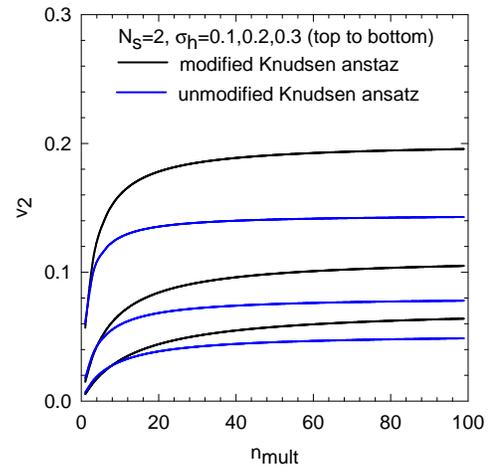}
}
\caption{(color online) 
Black and blue lines (from top to bottom) are predicted flow in modified and unmodified  Knudsen ansatz with two hot spots $N_s$=2, of sizes $\sigma_h$=0.1 fm.  
}
 \label{F4}
\end{figure}

  \begin{table}[h] 
\caption{\label{table1} Hydrodynamic limit $\vih$ and time and temperature scaled viscosity to entropy ratio from fit to PHOBOS data. The superscript ($*$) indicate that the vale was kept fixed during fitting.}
 \begin{tabular}{|c|c|c|c|}\hline
$K_0\sigma c_s (fm^2)$ &  $\vih$   & $\etas$ & $\chi^2/N$ \\
  \hline \hline
$0.05^*$ & 0.36 & 0.29 & 0.08\\ \hline
$0.23^*$ & 0.36 & 1.37 & 0.08\\ \hline 
$0.05^*$ & 0.26 & $0.0^*$ & 0.28\\ \hline
$0.23^*$ & 0.16 & $0.0^*$ & 2.20\\ \hline
\hline
\end{tabular} 
\end{table}

Knudsen ansatz predictions for elliptic flow in pp collisions are  shown in Fig.\ref{F3}. The blue lines are the predicted flow in the unmodified Knudsen ansatz, i.e. in the ideal fluid approximation, with two, three and four hot spots in the initial state. Hot spot size is assumed to be $\sigma_h$=0.1 fm. The predictions are obtained for $K_0\sigma c_s$=0.05. Depending on the number of hot spots, at large multiplicity predicted flow varies between 0.12-0.14. 
The black lines in Fig.\ref{F3} are the predicted flow in the modified Knudsen ansatz. The modified Knudsen ansatz predicts  $\sim$30\% more flow.  
We may note that if $K_0\sigma c_s$=0.23, instead of 0.05 is used, while the flow will remain unchanged in the modified Knudsen ansatz,
the unmodified Knudsen ansatz will predict $\sim$40\% less flow.

  \begin{table}[t] 
\caption{\label{table2} Minimum multiplicity ($n_{mult}^{min}$) beyond which modified Knudsen ansatz predictions for $v_2$ is accessible in 4th and 2nd order cumulant method are given as a function of hot spot numbers ($N_s$) and hot spot size $\sigma$ (in fm). The bracketed numbers are the
same in the unmodified Knudsen ansatz.}
 \begin{tabular}{|c|c|c|c||c|c|c|}\hline
 & \multicolumn{3}{c||}{$n_{mult}^{min}$ in 4th order cumulant} & \multicolumn{3}{c|}{$n_{mult}^{min}$ in 2nd order cumulant} \\ \hline
$N_s$& $\sigma=0.1$ &$\sigma=0.2$ &$\sigma=0.3$&$\sigma=0.1$&$\sigma=0.2$&$\sigma=0.3$\\
  \hline \hline
  \hline \hline  
2 & 12& 23& 48& 30& 76& $>100$\\ 
  & (15)& (29)& (63)& (51)& ($>100$)& ($>100$)\\ \hline
3 & 13& 25& 47& 32& 86& $>100$\\
 & (16)& (32)& (60)& (52)& ($>100$)& ($>100$)\\ \hline
4 & 15& 28& 48& 38& 96& $>100$\\ 
 & (18)& (35)& (62)& (63)& ($>100$)& ($>100$)\\ 
\hline
\end{tabular} 
\end{table} 

In Fig.\ref{F4}, Knudsen ansatz predictions for flow, as a function of the hot spot size are shown. Number of hot spots in the initial collisions  is assumed to be two. Predicted flow decreases with increasing hot spot size.  
For hot spot sizes 0.1, 0.2 and 0.3 fm,  
in the unmodified Knudsen ansatz, at large multiplicity, $v_2\sim $0.14, 0.08, 0.05 respectively. Modified Knudsen ansatz predicts $\sim$30\% more flow, $v_2\sim$ 0.2, 0.1, 0.06.

Is the predicted flow is sufficiently strong to be observed experimentally? As noted earlier, in 2nd and 4th order cumulant method, flow is measurable if $\vt \geq 1/n_{mult}^{1/2}$ and $\vf \geq 1/n_{mult}^{3/4}$. In table.\ref{table2}, the minimum multiplicity $n_{mult}^{min}$ beyond which Knudsen ansatz predicted flow become accessible in 4th and 2nd order cumulant method are noted. For 2-4 hot spots of size $\sigma_h$=0.1 fm, in 4th order cumulant method, modified Knudsen ansatz predicted flows are accessible beyond $n_{mult}^{min}$=12-15.
If hot spot sizes are large, flow is accessible only at larger multiplicity.
Limiting multiplicity is substantially 
larger in 2nd cumulant method, $n_{mult}^{min}$=30-38, for 2-4 hot spots of size 0.1 fm. Unmodified Knudsen ansatz predicts less flow and
demand higher multiplicity events for detection.

To summarise, assuming that     hot spot like structures are formed in pp collisions, in a modified Knudsen ansatz, which accounts for the entropy generation in viscous evolution, we have given predictions for the centrality dependence of elliptic flow in $\sqrt{s}$=14 TeV pp collision at LHC. 
Predicted flow depends on the number of hot spots as well as on the hot spot sizes. For 2-4 hot spots of size 0.1 fm, in large multiplicity events, modified Knudsen ansatz predicts $v_2\approx$0.18-0.20. Even in low multiplicity,
$n_{mult}\approx$ 10-15, events, predicted flow could be measured experimentally in 4th order  cumulant method.

\end{document}